\documentclass{article}
\usepackage{amssymb}
\usepackage{amsmath}
\usepackage[space]{cite}
\usepackage{tikz}
\usepackage{algorithm}
\usepackage{bbm}
\usepackage[noend]{algpseudocode}
\usepackage[normalem]{ulem}
\usepackage{url}
\usepackage{graphicx}
\usepackage{adjustbox}
\usepackage{caption}
\usepackage{subcaption}
\usepackage{enumitem}
\usepackage{xcolor}

\usepackage{MnSymbol}

\usepackage{cases}

\numberwithin{equation}{section}

\newcommand{\tr}{\mathop{\rm tr}\nolimits}
\newcommand{\kk}{\mathbbm{k}}
\newcommand{\atan}{\mathop{\rm atan2}\nolimits}

\usepackage{quantikz}
\usetikzlibrary{circuits.logic.US}
\tikzset{
    gateO/.style={
        draw,
        circle,
        minimum width=0.5em,
        inner sep=2pt    }
}
\DeclareExpandableDocumentCommand{\gateO}{O{}{m}}{|[gateO,#1]| {#2} \qw}

\tikzset{
    gateOS/.style={
        draw,
        circle,
        minimum width=0.5em,
        inner sep=2pt,
		fill=red!20}
}
\DeclareExpandableDocumentCommand{\gateOS}{O{}{m}}{|[gateOS,#1]| {#2} \qw}

\begin{document}

\begin{titlepage}
\vspace{0.5in}

\begin{center}

{\LARGE Spin-$s$ $U(1)$-eigenstate preparation}\\
\vspace{1in}

\large Nabi Zare Harofteh\footnote{\tt mxz866@miami.edu}\quad and 
\quad Rafael I. Nepomechie\footnote{\tt 
nepomechie@miami.edu}\renewcommand{\thefootnote}{*}\footnote{Corresponding author}
\renewcommand*{\thefootnote}{\arabic{footnote}}
\\[0.2in] 
Department of Physics, PO Box 248046\\[0.2in] 
University of Miami, Coral Gables, FL 33124 USA

\end{center}

\vspace{0.5in}

\begin{abstract}
We formulate a deterministic algorithm for preparing a general $U(1)$-eigenstate of a spin-$s$ chain of length $n$. These states consist of linear combinations of computational basis states $|\vec{m}\rangle$ of $n$ qudits, each with $(2s+1)$ levels and $s= 1/2, 1, 3/2, \ldots$, whose ditstrings $\vec{m}$ have a fixed digit sum. Exploiting a Gray code for bounded integer compositions, whose consecutive ditstrings obey the Gray property, the quantum state is prepared by applying corresponding ``Gray gates.'' We use this algorithm to prepare exact eigenstates of integrable spin-$s$ XXX Hamiltonians. We also consider the preparation of AKLT states and spin-$s$ Dicke states.
\end{abstract}

\end{titlepage}
 
\setcounter{footnote}{0}

\section{Introduction}

Many spin-chain Hamiltonians $H$ of physical interest have a $U(1)$ symmetry, e.g.
\begin{equation}
    \left[ H\,, \mathbb{S}^z \right] = 0 \,,
\end{equation}
where $\mathbb{S}^z = \sum_i S^z_i $ is the $z$-component of the total spin $\vec{\mathbb{S}}$, in which case one can look for simultaneous eigenstates $|\psi\rangle$ of $H$ and $\mathbb{S}^z$
\begin{align}
    H\, |\psi\rangle &= E\, |\psi\rangle \,, \nonumber \\
    \mathbb{S}^z\, |\psi\rangle &= M\, |\psi\rangle \,.
\end{align}
The problem of preparing such states on a quantum computer has drawn considerable attention, see e.g. \cite{Bartschi2019, Wang:2021, Buhrman:2023rft, Bond:2023yry, Piroli:2024ckr, Yu:2024szp, Liu:2024taj, Vasconcelos:2026sqs, VanDyke:2021kvq, VanDyke:2021nuz,Li:2022czv,Sopena:2022ntq, Ruiz:2023rew, Ruiz:2025qmt, Sahu:2024toz, Yeo:2025tph, Raveh:2024llj, Farias:2024ejh,Mao:2024hfg, Li:2025zzw, Luo:2025zjt, Smith:2022nbd, Kumaran:2024atr, Nepomechie:2024fhm, Raveh:2024sku, Kerzner:2025uxw}.

In the case of a spin-1/2 chain of length $n$ (equivalently, a system of $n$ qubits), the $\mathbb{S}^z$-eigenstates consist of linear combinations of computational basis states $|m_n \ldots m_2\, m_1\rangle$, $m_i  \in \{0, 1\}$ for all $i$, with a fixed Hamming weight\footnote{Here $|0\rangle=\binom{1}{0}\,, |1\rangle=\binom{0}{1}$, and tensor products are understood $|m_n \ldots m_2\, m_1\rangle = |m_n \rangle \otimes\ldots \otimes |m_2\rangle \otimes |m_1\rangle$.}
\begin{equation}
|\psi\rangle = \sum_{\substack{m_i = 0, 1 \\
m_1 + m_2 + \cdots + m_n = k}} 
a_{\vec{m}}
\; |m_n \ldots m_2\, m_1\rangle \,, 
\label{qubitsup}
\end{equation}
where the Hamming weight $k \in \{ 0, 1, \ldots,  n\} $ is related to the $\mathbb{S}^z$-eigenvalue $M$ by $k=\frac{n}{2}-M$.
Deterministic algorithms for preparing such states with generic normalized amplitudes $a_{\vec{m}}$ were first presented in \cite{Raveh:2024llj} and 
\cite{Farias:2024ejh}. The former algorithm, which focuses on Bethe states (eigenstates of integrable Hamiltonians, such as XXZ, whose amplitudes $a_{\vec{m}}$ are given by the coordinate Bethe ansatz \cite{Bethe:1931hc, Gaudin:1983, Faddeev:1996iy}), exploits recursion \cite{Bartschi2019}; while the latter algorithm exploits the fact that the list of bitstrings\footnote{We denote bitstrings (and, more generally, ditstrings) interchangeably by $m_n \ldots m_2\, m_1$ and $\vec{m}=(m_n, \ldots, m_2, m_1)$.} $\vec{m} = (m_n, \ldots, m_2, m_1)$ in the superposition \eqref{qubitsup} can be ordered in such a way that consecutive bitstrings $\vec{m}$ and $\vec{m'}$ satisfy the so-called Gray property
\begin{equation}
    \vec{m'} = \vec{m} + \hat{e}_i - \hat{e}_j \,,
    \label{Gray}
\end{equation}
where $i, j \in \{1, \ldots n\}$, $i \ne j$,  and $\hat{e}_i$ is the $n$-dimensional
unit vector with components $(\hat{e}_i)_l = \delta_{i,l}$. In other words, $\vec{m'}$ and $\vec{m}$ differ only at positions $i$ and $j$, where the bit increases (decreases) by 1, respectively.
A list of bitstrings satisfying 
\eqref{Gray} is called a Gray code, see e.g. \cite{Wilf:1989, savage:1997, mutze:2012}. Additional approaches for preparing generic states in a fixed Hamming-weight subspace are presented in \cite{Mao:2024hfg, Li:2025zzw, Luo:2025zjt}.
Alternative approaches for preparing 
spin-1/2 XXZ Bethe states are discussed in \cite{VanDyke:2021kvq, VanDyke:2021nuz, Li:2022czv, Sopena:2022ntq, Ruiz:2023rew, Ruiz:2025qmt, Sahu:2024toz, Yeo:2025tph}.
The states \eqref{qubitsup} with $a_{\vec{m}} = 1/\sqrt{\binom{n}{k}}$ (independent of $\vec{m}$) are the so-called Dicke states \cite{Dicke:1954zz}, for which more efficient preparation schemes are available, see e.g. \cite{Bartschi2019, Wang:2021, Buhrman:2023rft, Bond:2023yry, Piroli:2024ckr, Yu:2024szp, Liu:2024taj, Vasconcelos:2026sqs}.

The main goal of this paper is to deterministically prepare generic normalized $\mathbb{S}^z$-eigenstates of a spin-$s$ chain of length $n$ (equivalently, a system of $n$ $d$-level qudits, where $d=2s+1$)
\begin{equation}
|\psi\rangle = \sum_{\substack{m_i = 0, 1, \dots, 2s \\
m_1 + m_2 + \cdots + m_n = k}} 
a_{\vec{m}}
\; |m_n \ldots m_2\, m_1\rangle \,,
\label{quditsup}
\end{equation}
where $s\in \{1/2, 1, 3/2, \ldots \}$, and 
$k \in \{ 0, 1, \ldots,  2 s n\} $ is related to the $\mathbb{S}^z$-eigenvalue $M$ by $k=s\, n -M$.
For $s>1/2$, the computational basis states $|\vec{m}\rangle$ in \eqref{quditsup} do {\em not} have fixed Hamming weight, which is defined for multi-qudit states
as the number of nonzero $m_i$'s; instead, these states are characterized by the {\em fixed digit sum} $k$. Remarkably, as in the case $s=1/2$ considered in \cite{Farias:2024ejh}, it is always possible to order the list of ``ditstrings'' $\vec{m}$ in the superposition \eqref{quditsup} such that consecutive ditstrings have the Gray property \eqref{Gray}. We exploit this fact here to design a quantum circuit that builds up the superposition of states \eqref{quditsup} one state at a time, by successively rotating $|\vec{m}\rangle$ to a linear combination of $|\vec{m}\rangle$ and $|\vec{m'}\rangle$. The quantum circuit makes use of Gray gates \eqref{Ggate} that generally have multiple controls; the number of such gates needed to prepare the
most general state \eqref{quditsup} is $\mathcal{O(D)}$ \eqref{dim}, which in the 
worst case $k=\lfloor s n\rfloor$ scales exponentially with $n$.

The remainder of this paper is organized as follows. In Sec. \ref{sec:algorithm}, we present our algorithm for preparing the state \eqref{quditsup}. 
As applications of this result, we consider
in Sec. \ref{sec:AKLT} the preparation of the ground state of the AKLT model \cite{Affleck:1987vf, Affleck:1987cy}, and 
in Sec. \ref{sec:Dicke} the preparation of spin-$s$ Dicke states \cite{Nepomechie:2024fhm, Raveh:2024sku}.
In Sec. \ref{sec:Bethe}, we discuss the preparation of exact eigenstates of 
spin-$s$ integrable XXX Hamiltonians \cite{Zamolodchikov:1980ku, Kulish:1981gi, Kulish:1981bi, Babujian:1983ae, Lima:1999, Crampe:2011}.
While more efficient preparation schemes are available for both AKLT states and spin-$s$ Dicke states (see \cite{Smith:2022nbd, Kumaran:2024atr} and \cite{Nepomechie:2024fhm, Kerzner:2025uxw}, respectively),
we are not aware of alternative proposals for preparing spin-$s$ Bethe states. We conclude 
in Sec. \ref{sec:Discussion} with a brief discussion of potential additional applications and directions for further investigation. 
Implementations in cirq \cite{cirq} of the algorithm and its applications are available on GitHub \cite{GitHubRN}.

\section{General algorithm}\label{sec:algorithm}

In this section, we present our algorithm for preparing the state \eqref{quditsup}.
Before entering into details, let us illustrate the basic idea of
how this algorithm works for the case $(n,k,s) = (3, 3, 1)$
with real amplitudes $a_{\vec{m}}$. We begin by observing that, restricting to digits 0,1,2 (since here $s=1$), the following list of $n=3$-digit ditstrings, with each ditstring having fixed digit sum $k=3$,
\begin{equation}
    012, 021, 120, 111, 102, 201, 210
    \label{code-example}
\end{equation}
is a Gray code: consecutive ditstrings have the Gray property \eqref{Gray}, see also Table \ref{table:GrayExample}. 

\begin{table}[ht]
\centering
\begin{tabular}{|c|c|c|c|c|c|}
\hline
$l$ & $\vec{m}^{[l]}$ & $i$ & $j$ & $m_i^{[l]}$ & $m_j^{[l]}$ \\   
\hline
0 & 0\textcolor{red}{1}\textcolor{blue}{2}  & 2 & 1 & \textcolor{red}{1} & \textcolor{blue}{2} \\
1 & \textcolor{red}{0}2\textcolor{blue}{1}  & 3 & 1 & \textcolor{red}{0} & \textcolor{blue}{1} \\
2 & 1\textcolor{blue}{2}\textcolor{red}{0}  & 1 & 2 & \textcolor{red}{0} & \textcolor{blue}{2} \\
3 & 1\textcolor{blue}{1}\textcolor{red}{1}  & 1 & 2 & \textcolor{red}{1} & \textcolor{blue}{1} \\
4 & \textcolor{red}{1}0\textcolor{blue}{2}  & 3 & 1 & \textcolor{red}{1} & \textcolor{blue}{2} \\
5 & 2\textcolor{red}{0}\textcolor{blue}{1}  & 2 & 1 & \textcolor{red}{0} & \textcolor{blue}{1} \\
6 & 210  &   &   &  & \\
\hline
\end{tabular}
\caption{A Gray code of ditstrings $\vec{m}^{[l]}$ for $(n,k,s) = (3, 3, 1)$; the $(i,j)$ values such that $\vec{m}^{[l+1]} = \vec{m}^{[l]} + \hat{e}_i -\hat{e}_j$; and the corresponding values $(m_i^{[l]} ,m_j^{[l]})$. The digits in the ditstrings are labeled from right (1) to left (3).}
\label{table:GrayExample}
\end{table}

The quantum circuit begins by preparing the state $|012\rangle$ (corresponding to the leftmost element in \eqref{code-example}), and then applying successive transformations to the basis states
\begin{equation}
|\vec{m}\rangle \rightarrow \cos(\theta)\, |\vec{m}\rangle + \sin(\theta)\, |\vec{m'}\rangle
\label{rotation}
\end{equation}
in the order \eqref{code-example} read left-to-right by angles $\theta_0, \ldots, \theta_5$:
\begin{align}
    |012\rangle &\rightarrow c_0 |012\rangle + s_0 |021\rangle \nonumber\\
  &\rightarrow c_0 |012\rangle + s_0 ( c_1 |021\rangle + s_1 |120\rangle) \nonumber\\  
  &\rightarrow c_0 |012\rangle + s_0 c_1 |021\rangle + s_0 s_1 (c_2 |120\rangle +  s_2 |111 \rangle) \nonumber\\
  &\rightarrow c_0 |012\rangle + s_0 c_1 |021\rangle + s_0 s_1 c_2 |120\rangle 
  +  s_0 s_1 s_2 (c_3 |111 \rangle + s_3 |102 \rangle) \nonumber\\
  &\rightarrow c_0 |012\rangle + s_0 c_1 |021\rangle + s_0 s_1 c_2 |120\rangle 
  +  s_0 s_1 s_2 c_3 |111 \rangle + s_0 s_1 s_2 s_3 (c_4 |102 \rangle + s_4 |201 \rangle) \nonumber\\
  &\rightarrow c_0 |012\rangle + s_0 c_1 |021\rangle + s_0 s_1 c_2 |120\rangle 
  +  s_0 s_1 s_2 c_3 |111 \rangle + s_0 s_1 s_2 s_3 c_4 |102 \rangle 
  + s_0 s_1 s_2 s_3 s_4 (c_5 |201 \rangle + s_5 |210 \rangle) \nonumber\\
  &= c_0 |012\rangle + s_0 c_1 |021\rangle + s_0 s_1 c_2 |120\rangle 
  +  s_0 s_1 s_2 c_3 |111 \rangle + s_0 s_1 s_2 s_3 c_4 |102 \rangle 
  + s_0 s_1 s_2 s_3 s_4 c_5 |201 \rangle + s_0 s_1 s_2 s_3 s_4 s_5 |210 \rangle \,,
  \label{state-example}
\end{align}
where $c_l = \cos(\theta_l)\,, s_l = \sin(\theta_l)$, and the angles are chosen
so that the amplitudes in the final state \eqref{state-example} match with the amplitudes $a_{\vec{m}}$ in the given state \eqref{quditsup}, see \eqref{need-real}, \eqref{thetas-real}.

As can be seen from this example, the two key ingredients in the algorithm are the Gray code \eqref{code-example} and the corresponding transformations \eqref{rotation}, which we discuss in more detail in Secs. \ref{sec:Gray} and \ref{sec:rotation}, respectively. The full algorithm is presented in Sec. \ref{sec:full}.

\subsection{Gray code for bounded integer compositions}\label{sec:Gray}

In the mathematics literature, the ditstrings $\vec{m}$ corresponding to the basis states  $|\vec{m}\rangle$ in \eqref{quditsup} are known as ``bounded integer compositions,'' since $k$ is composed into $n$ parts $m_1, \ldots, m_n$ that are bounded $ 0 \le m_i \le 2s$ for all $i$. Remarkably, it has been shown by Walsh \cite{Walsh:2000} (based on earlier work by Knuth \cite{Wilf:1989} and Klingsberg \cite{Klingsberg:1982} for unrestricted integer compositions)
that a Gray code with the property \eqref{Gray} exists for any valid values of $(n,k,s)$.\footnote{The work \cite{Farias:2024ejh}, which is restricted to bitstrings with fixed Hamming weight, makes use of a simpler Gray code due to Ehrlich \cite{Even:1973}.}

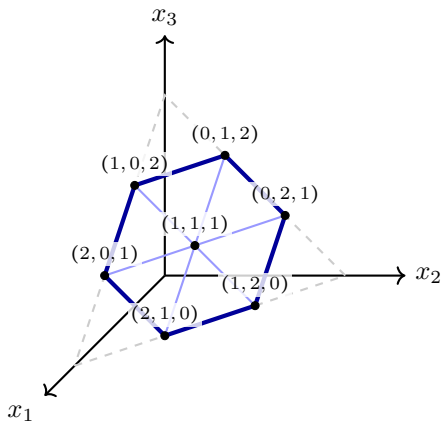
\begin{figure}[htbp] 
    \centering 
\begin{tikzpicture}[x={(-0.4cm,-0.4cm)}, y={(0.8cm,0cm)}, z={(0cm,0.8cm)}, scale=1.0]

    \draw[->, thick] (0,0,0) -- (4,0,0) node[below left] {$x_1$};
    \draw[->, thick] (0,0,0) -- (0,4,0) node[right] {$x_2$};
    \draw[->, thick] (0,0,0) -- (0,0,4) node[above] {$x_3$};

    \draw[gray!40, thick, dashed] (3,0,0) -- (0,3,0) -- (0,0,3) -- cycle;

    \coordinate (A) at (2,1,0);
    \coordinate (B) at (1,2,0);
    \coordinate (C) at (0,2,1);
    \coordinate (D) at (0,1,2);
    \coordinate (E) at (1,0,2);
    \coordinate (F) at (2,0,1);
    \coordinate (Center) at (1,1,1);

    \draw[ultra thick, blue!60!black] (A) -- (B) -- (C) -- (D) -- (E) -- (F) -- cycle;

    \draw[thick, blue!40] (Center) -- (A);
    \draw[thick, blue!40] (Center) -- (B);
    \draw[thick, blue!40] (Center) -- (C);
    \draw[thick, blue!40] (Center) -- (D);
    \draw[thick, blue!40] (Center) -- (E);
    \draw[thick, blue!40] (Center) -- (F);

    \foreach \p/\l in {A/{(2,1,0)}, B/{(1,2,0)}, C/{(0,2,1)}, D/{(0,1,2)}, E/{(1,0,2)}, F/{(2,0,1)}, Center/{(1,1,1)}}
    {
        \filldraw[black] (\p) circle (1.5pt);
        \node[fill=white, opacity=0.8, text opacity=1, inner sep=1pt, font=\scriptsize] at (\p) [yshift=8pt] {$\l$};
    }

\end{tikzpicture}
\caption{The convex polytope for bounded compositions of $k=3$ into $n=3$ parts with $0 \le x_i \le 2$. The hexagon is the intersection of the simplex $x_1+x_2+x_3=3$ and the cube $0 \le x_i \le 2$.}
    \label{fig:hexagon}
\end{figure}

Walsh \cite{Walsh:2000} found an algorithm, reviewed in Appendix \ref{sec:Walsh}, that can be used to generate the needed Gray code, 
see also \cite{horan:2014}.
Alternatively, for $n<20$, the Gray code can be generated quickly by searching on the associated graph for a Hamiltonian path (i.e., a path that visits every vertex exactly once) \cite{mutze:2012}.
A set of bounded integer compositions can be associated with the integer lattice points contained within or on the boundary of a corresponding convex polytope. For the example $n=3, k=3, s=1$ given by \eqref{code-example}, the corresponding convex polytope is the regular hexagon shown in Fig. \ref{fig:hexagon}. These lattice points can be regarded as vertices of a graph; and an edge exists between vertices $\vec{m}$ and $\vec{m'}$ if their Manhattan distance, defined by
\begin{equation}
    d(\vec{m}\,, \vec{m'}) = \sum_{i=1}^n |m_i - m'_i| \,,
\end{equation}
is exactly 2, see \eqref{Gray}. Hence, finding a Gray code is equivalent to finding a Hamiltonian path on the associated graph.

To this end, one can perform a backtracking search using Warnsdorff's rule,
see e.g. \cite{GitHubRN}:
\begin{enumerate}
    \item Start at a vertex
    \item Find all its unvisited neighbors (distance=2)
    \item Pick the neighbor that has the {\em fewest} remaining unvisited neighbors (Warnsdorff's rule)
    \item Move and repeat; if reach a dead end (unable to visit all the vertices), backtrack and try the next-best neighbor.
\end{enumerate}

\subsection{Gray gate}\label{sec:rotation}

We now construct a quantum ``Gray gate'' $G_{i,j}^{m_i,m_j}(\theta,\phi)$ that performs the following unitary transformation on the computational basis states $|\mu \rangle_i\,  |\nu \rangle_j$ with $\mu, \nu  \in \{0, 1, \ldots, 2s\}$
of two $d=(2s+1)$-level qudits labeled by $i, j \in \{1, 2, \ldots, n\}$, $i \ne j$, 
\begin{equation}
G_{i,j}^{m_i,m_j}(\theta,\phi)\, \left( |\mu \rangle_i\,  |\nu \rangle_j \right) =
\begin{cases}
\cos(\theta)\,  |m_i \rangle_i\,  |m_j \rangle_j
+ e^{i \phi}  \sin(\theta)\,    |m_i +1 \rangle_i\,  |m_j -1 \rangle_j 
& (\mu, \nu) = (m_i, m_j) \\
e^{i \phi} \cos(\theta)\,  |m_i +1 \rangle_i\,  |m_j -1 \rangle_j
 -  \sin(\theta)\,  |m_i \rangle_i\,  |m_j \rangle_j
& (\mu, \nu) = (m_i+1, m_j-1) \\
|\mu \rangle_i\,  |\nu \rangle_j 
& (\mu, \nu) \ne (m_i, m_j)\,, (m_i+1, m_j-1)
\end{cases} 
\label{Ggate}
\end{equation}
for given values of $m_i, m_j \in \{0, 1, \ldots, 2s\}$. From the {\em first} line of \eqref{Ggate}, we see that this gate performs (when augmented with suitable controls on other qudits, see Sec. \ref{sec:controls}) the desired $n$-qudit transformation
\begin{equation}
 |\vec{m}\rangle \rightarrow
 \cos(\theta)\,  |\vec{m}\rangle
+ e^{i \phi}  \sin(\theta)\,    |\vec{m'}\rangle \,,
\label{Graytransf}
\end{equation}
where $\vec{m'}$ is related to $\vec{m}$ by the Gray property \eqref{Gray}. Note that the transformation \eqref{Ggate} is not symmetric in $i$ and $j$.

\begin{figure}[htb]
	\centering
\begin{adjustbox}{width=0.6\textwidth}
\begin{quantikz}
\lstick{$j$} & \gateO{m_j} \vqw{1} & \gate{R_{m_j,m_j-1}(\theta,\phi)} \vqw{1} 
& \gateO{m_j} \vqw{1} & \qw & \qw \rstick[2, brackets=none]{$\quad =\quad$}\\
\lstick{$i$} & \gate{X}  & \gateO{m_i+1} & \gate{X^\dagger}  & \qw & \qw \\
\end{quantikz}
\begin{quantikz}
& \gate[label style={label={above:$\theta, \phi$}}]{\flat\ m_j} \vqw{1} & \qw \\
& \gate{\sharp\ m_i} & \qw \\
\end{quantikz}
\end{adjustbox}
\caption{Circuit diagram for the Gray gate $G_{i,j}^{m_i,m_j}(\theta,\phi)$ \eqref{Ggate}. In its symbolic form shown on the right, we use sharp ($\sharp$) and flat ($\flat$) to designate qudits $i$ and $j$, respectively.}
\label{fig:Ggate}
\end{figure}
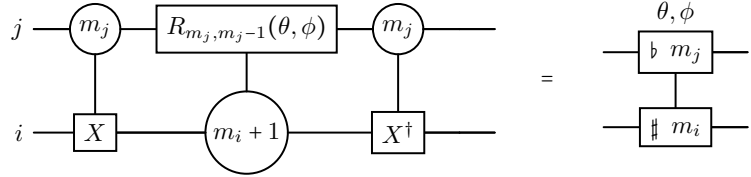

The transformation $G_{i,j}^{m_i,m_j}(\theta,\phi)$ \eqref{Ggate} can be implemented with the quantum circuit shown in Fig. \ref{fig:Ggate}.

\noindent
The horizontal wires represent the two $d$-level qudits.  A circle $\begin{quantikz}\gateO{\scriptscriptstyle \mu}\end{quantikz}$ denotes a control on the value $\mu$. 
The 1-qudit shift gate ${\rm X}$ and its inverse are defined as (see e.g. \cite{Wang:2020})
\begin{equation}
{\rm X}\, |\mu\rangle = |\mu+1 \rangle\,, \qquad 
{\rm X}^\dagger\, |\mu\rangle = |\mu-1 \rangle \,,
\label{Xgate}
\end{equation}
where the sums are defined modulo $d$; these gates are controlled by the upper wire in Fig. \ref{fig:Ggate}. Moreover, $R_{m_j,m_j-1}(\theta,\phi)$ is the 1-qudit unitary gate 
(Givens rotation) defined by
\begin{align}
R_{m_j,m_j-1}(\theta,\phi) &= \cos(\theta) |m_j\rangle\langle m_j | 
-  \sin(\theta) |m_j\rangle\langle m_j-1 |   \nonumber \\
&+ e^{i \phi} \sin(\theta) |m_j-1\rangle\langle m_j | 
+ e^{i \phi} \cos(\theta) |m_j-1\rangle\langle m_j-1 | 
+ \sum_{\mu \ne m_j\,, m_j-1} |\mu\rangle\langle \mu|\,,
\end{align}
which in Fig. \ref{fig:Ggate} is controlled by the lower wire. Our Gray gate $G_{i,j}^{m_i,m_j}(\theta,\phi)$ is a qudit generalization of the qubit RBS gate in \cite{Farias:2024ejh}, which is similar to the I gate in \cite{Bartschi2019, Raveh:2024llj}; similar qudit gates appear in \cite{Nepomechie:2024fhm, Raveh:2024sku}.

\subsubsection{Controls on other qudits}\label{sec:controls}

The Gray transformations \eqref{Graytransf} are generally applied in succession, as illustrated in \eqref{state-example}. In order to avoid transforming previously-generated states, it is generally necessary to add to the Gray gates $G_{i,j}^{m_i,m_j}$ controls on qudits other than $i, j$. We now proceed to determine the labels (i.e., addresses, or indices) of these other qudits, and the values of the controls. 

To this end, let us label the ditstrings in the Gray code by $\vec{m}^{[l]}\,, l = 0, 1, \ldots$, as in Table \ref{table:GrayExample}.
Let us recall from \eqref{Gray} that 
\begin{equation}
\vec{m}^{[l+1]} - \vec{m}^{[l]} = \hat{e}_{i^{[l]}} -\hat{e}_{j^{[l]}} \,,
\label{Grayagain}
\end{equation}
where we have now appended $[l]$ to $i$ and $j$ to avoid ambiguity.
That is, all the components of $\vec{m}^{[l+1]}$ and $\vec{m}^{[l]}$ are equal except for those labeled $i^{[l]}$ and $j^{[l]}$. The set difference\footnote{By definition, the set difference $A\backslash B$ is the set of elements in set $A$ that are not in set $B$.} 
\begin{equation}
S^{[l]}:=\{1, \ldots, n\}\backslash \{i^{[l]}, j^{[l]}\}
\end{equation}
is therefore the set of labels of components of $\vec{m}^{[l+1]}$ and $\vec{m}^{[l]}$ that are {\em equal}. Let $C^{[l]} \subset S^{[l]}$ be the set of labels of components of $\vec{m}^{[l+1]}$ and $\vec{m}^{[l]}$ that are {\em equal and nonzero}; that is
\begin{equation}
C^{[l]} := \{ r \in S^{[l]}: m_r^{[l+1]} = m_r^{[l]} \ne 0 \} \,.
\label{setC}
\end{equation}
Naively, a control should be placed at all $r \in C^{[l]}$, with control value $m_r^{[l+1]} = m_r^{[l]}$.

However, some of these controls are redundant, and can be safely removed. The unnecessary controls can be pruned by the following generalization of Subroutine 2 in \cite{Farias:2024ejh}:
Let $U^{[0]}$ be the set of labels of the nonzero ``untouched'' components of the initial ditstring $\vec{m}^{[0]}$; that is, 
\begin{equation}
U^{[0]} := \{ r  \in \{1, \ldots, n\} :\ m^{[0]}_r  \ne 0 \} \,.
\label{setU0}
\end{equation}
At step $l$, define $C^{[l]}$ as before \eqref{setC}; that is
\begin{equation}
C^{[l]} := \{ r \in \{1, \ldots, n\}\backslash \{i^{[l]}, j^{[l]}\}:\ m_r^{[l+1]} = m_r^{[l]} \ne 0 \} \,.
\label{setCagain}
\end{equation}
Set 
\begin{equation}
U^{[l]} =  U^{[l-1]}\backslash \{i^{[l]}, j^{[l]}\} \,, \qquad l>0\,,
\end{equation}
and use it to prune $C^{[l]}$
\begin{equation}
C^{[l]} := C^{[l]}\backslash U^{[l]} \,.
\label{setCpruned}
\end{equation}
Again, a control should be placed at all $r \in C^{[l]}$, with control value $m_r^{[l+1]} = m_r^{[l]}$. These controls should be added to the (single-controlled) Givens rotation in Fig. \ref{fig:Ggate}.
The maximum number of controls of a Gray gate is ${\rm min}(n-2,k-1)$.

\subsection{Full algorithm}\label{sec:full}

We now present our full algorithm for preparing the state \eqref{quditsup} for given values of $(n,k,s)$ and specified amplitudes $a_{\vec{m}}$. (In the examples of Secs. \ref{sec:AKLT},  \ref{sec:Dicke}, \ref{sec:Bethe}, the amplitudes $a_{\vec{m}}$ are given functions of $\vec{m}$, see Eqs. \eqref{AKLT-MPS}, \eqref{Dicke-amplitudes}, \eqref{Bethe-amplitudes}, respectively.) The number of terms $\mathcal{D}$ in the sum \eqref{quditsup} can be shown to be given by
\begin{equation}
\mathcal{D} = \sum_{l=0}^{\lfloor k/(2s+1) \rfloor}(-1)^l \binom{n}{l} \binom{k-l(2s+1)+n-1}{n-1} \,,
\label{dim}
\end{equation}
which simply reduces to $\binom{n}{k}$ for $s=1/2$. 

\begin{enumerate}

\item The Gray code for $(n,k,s)$ is generated classically, e.g. using the Hamiltonian search described in Section \ref{sec:Gray}, or Walsh's algorithm reviewed in Appendix \ref{sec:Walsh}. As before, the ditstrings are labeled $\vec{m}^{[l]}\,, l = 0, 1, \ldots, \mathcal{D}-1$, where $\mathcal{D}$ is given by \eqref{dim}. 

\item The values $(i^{[l]}, j^{[l]})$, $(m_{i^{[l]}}^{[l]}, m_{j^{[l]}}^{[l]})$ and angles $(\theta_l, \phi_l)$ that parametrize the Gray gates \eqref{Ggate}, as well as the labels and values of the controls, are computed classically. The values  $(i^{[l]}, j^{[l]})$, and therefore $(m_{i^{[l]}}^{[l]}, m_{j^{[l]}}^{[l]})$, follow immediately from \eqref{Grayagain}. The labels and values of the controls are computed from \eqref{setU0}-\eqref{setCpruned}. To compute the angles, it is convenient to denote the amplitudes now by $a_l := a_{\vec{m}^{[l]}}$. 

For the case of all real amplitudes, we can restrict to transformations \eqref{Graytransf} with $\phi=0$, as in \eqref{rotation}. We therefore set 
\begin{equation}
a_l = \begin{cases}
    \left(\prod_{j=0}^{l-1} \sin(\theta_j) \right) \cos(\theta_l) & l = 0, \ldots, \mathcal{D}-2 \\[0.1 in]
        \prod_{j=0}^{\mathcal{D}-2} \sin(\theta_j) & l = \mathcal{D}-1
    \end{cases} \,,
\label{need-real}
\end{equation}
as in \eqref{state-example}. It follows that, similarly to \cite{Farias:2024ejh},
the angles $\theta_l$ are given by 
\begin{equation}
\theta_l = \begin{cases}
\atan \left( \sqrt{\sum_{j=l+1}^{\mathcal{D}-1} a_j^2}, a_l \right) &  l = 0, \ldots, \mathcal{D}-3 \\[0.1 in]
\atan \left(a_{\mathcal{D}-1}, a_{\mathcal{D}-2} \right) & l = \mathcal{D}-2
    \end{cases} \,,
\label{thetas-real}
\end{equation}
where $\atan$ is the 2-argument arctangent function, such that $\atan(y,x)={\rm Arg}(x+i y) \in (-\pi, \pi]$.

For the case of complex amplitudes, we perform transformations \eqref{Graytransf} with nonzero $\phi$. Hence, instead of \eqref{need-real}, we set\footnote{If necessary, all the amplitudes should be rescaled $a_l \mapsto a_l\, |a_0|/a_0$ so that $a_0$ is real.}
\begin{equation}
a_l = \begin{cases}
    \left(\prod_{j=0}^{l-1} e^{i \phi_j} \sin(\theta_j) \right) \cos(\theta_l) & l = 0, \ldots, \mathcal{D}-2 \\[0.1 in]
        \prod_{j=0}^{\mathcal{D}-2}  e^{i \phi_j} \sin(\theta_j) & l = \mathcal{D}-1
    \end{cases} \,.
\end{equation}
Therefore, the angles $\theta_l, \phi_l$ are given by
\begin{equation}
\theta_l = \begin{cases}
\atan \left( \sqrt{\sum_{j=l+1}^{\mathcal{D}-1} |a_j|^2}, |a_l| \right) &  l = 0, \ldots, \mathcal{D}-3 \\[0.1 in]
\atan \left(|a_{\mathcal{D}-1}|, |a_{\mathcal{D}-2}| \right) & l = \mathcal{D}-2
    \end{cases} \,,
\label{thetas-complex}
\end{equation}
and
\begin{equation}
\phi_l = 
\arg (a_{l+1}) - \arg  (a_l)\,, \qquad l = 0, \ldots, \mathcal{D}-2 \,.
\label{phis-complex}
\end{equation}

\item The quantum part of the algorithm begins by preparing the state $|\vec{m}^{[0]}\rangle$, which is achieved by applying corresponding powers of
${\rm X}$ gates \eqref{Xgate} to the all-$|0\rangle$ state
\begin{equation}
|\vec{m}^{[0]}\rangle = \prod_{r=1}^n
{\rm X}^{m^{[0]}_{r}}_r\, |0\rangle^{\otimes n} \,.
\end{equation}

\item Finally, the Gray gates \eqref{Ggate}, controlled according to \eqref{setU0}-\eqref{setCpruned}, are applied successively to prepare the state \eqref{quditsup}
\begin{equation}
|\psi\rangle = \overset{\curvearrowleft}{\prod_{l=0}^{\mathcal{D}-2}} G^{[l]} \, |\vec{m}^{[0]}\rangle \,,
\label{genalgorithm}
\end{equation}
where $G^{[l]} = 
G_{i^{[l]},j^{[l]}}^{m^{[l]}_{i^{[l]}},m^{[l]}_{j^{[l]}}}(\theta_l,\phi_l)$,
and the product goes from right to left with increasing $l$. Note that no ancillary qudits are needed.
\end{enumerate}

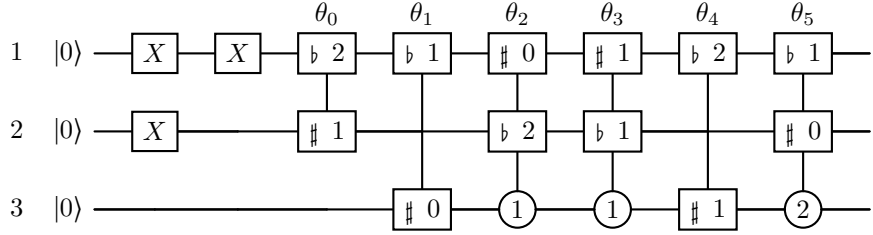
\begin{figure}[htb]
	\centering
\begin{adjustbox}{width=0.7\textwidth}
\begin{quantikz}
\lstick{$1\quad \ket{0}$} & \gate{X} & \gate{X}
& \gate[label style={label={above:$\theta_0$}}]{\flat\ 2} \vqw{1} & \gate[label style={label={above:$\theta_1$}}]{\flat\ 1} \vqw{2}  
& \gate[label style={label={above:$\theta_2$}}]{\sharp\ 0} \vqw{1} 
&  \gate[label style={label={above:$\theta_3$}}]{\sharp\ 1} \vqw{1}  
& \gate[label style={label={above:$\theta_4$}}]{\flat\ 2} \vqw{2} 
& \gate[label style={label={above:$\theta_5$}}]{\flat\ 1} \vqw{1} & \qw\\
\lstick{$2\quad \ket{0}$} & \gate{X} & \qw & \gate{\sharp\ 1} 
& \qw  & \gate{\flat\ 2} \vqw{1} & \gate{\flat\ 1} \vqw{1}
& \qw  & \gate{\sharp\ 0}  \vqw{1}  & \qw\\
\lstick{$3\quad \ket{0}$} & \qw  & \qw  & \qw & \gate{\sharp\ 0} 
&  \gateO{1} &  \gateO{1} & \gate{\sharp\ 1}  &  \gateO{2} & \qw\\
\end{quantikz}
\end{adjustbox}
\caption{The quantum circuit \eqref{genalgorithm} for preparing a state with $(n,k,s) = (3, 3, 1)$, see \eqref{state-example} and Table \ref{table:GrayExample}. The Gray gates are defined in Fig. \ref{fig:Ggate}, and the circles denote controls.}
\label{fig:CircuitExample}
\end{figure}

For the example \eqref{state-example} with $(n,k,s) = (3, 3, 1)$, the quantum circuit is shown in Fig. \ref{fig:CircuitExample}.

The size and depth of the quantum circuit \eqref{genalgorithm} is evidently $\mathcal{O}(\mathcal{D})$, where $\mathcal{D}$ is given by \eqref{dim}. For $n\gg k$ and $s \ge 1/2$, the $l=0$ term dominates the sum, hence $\mathcal{D}$ scales polynomially with $n$:
$\mathcal{D} \sim \binom{n+k-1}{k} \sim \frac{n^k}{k!}$. For the worst case $k=\lfloor s n\rfloor$,  $\mathcal{D}$ scales exponentially with $n$: $\mathcal{D} \sim \frac{(2s+1)^n}{\sqrt{2\pi n s (s+1)/3}}$, and the maximum number of controls per Gray gate is $\mathcal{O}(n)$. 

Refining the estimate for the scaling of the quantum circuit depth in terms of numbers of two-qudit gates, we recall \cite{Wang:2020, Zi:2023drn} that $\ell$-controlled qudit gates can be synthesized using $\mathcal{O}(\ell)$ two-qudit gates with one ancilla.  The total number of two-qudit gates therefore scales approximately as $\mathcal{O}(\ell\, \mathcal{D})$, where $\ell= {\rm min}(n-2,k-1)$ is the maximum number of controls of a Gray gate.

This qudit-based algorithm can be implemented using
qubits, albeit with a significant compilation overhead.  Indeed, each
$d=(2s+1)$-level qudit can be mapped to qubits using e.g. a binary
encoding (each qudit is mapped to $\log_{2}(2s+1)$ qubits; for
example, $s=1$ requires 2 qubits, such that the single-qudit states
$|0\rangle, |1\rangle, |2\rangle$ are mapped to the 2-qubit states
$|00\rangle, |01\rangle, |10\rangle$, respectively) or unary encoding
(each qudit is mapped to $2s$ qubits according to the Hamming weight
of the latter; for example, $s=3/2$ requires 3 qubits, such that the
single-qudit states $|0\rangle, |1\rangle, |2\rangle, |3\rangle$ are
mapped to the 3-qubit states $|000\rangle, |001\rangle, |011\rangle,
|111\rangle$, respectively).  The single-qudit shift ($X$) and
rotation ($R$) gates in a Gray gate will translate to multi-controlled
qubit gates; and each single-qudit control will translate to a multi-qubit
control.  Hence, implementing the algorithm using qubits will
significantly increase the circuit size and depth. A general 
discussion of the advantage of using qudits compared with qubits can be 
found in \cite{Wang:2020}.

\section{AKLT ground state}\label{sec:AKLT}

The AKLT Hamiltonian \cite{Affleck:1987vf, Affleck:1987cy} is given by
\begin{equation}
    H=\sum_{i=1}^n \left(\frac{1}{2} \vec{S}_i \cdot \vec{S}_{i+1} + \frac{1}{6} (\vec{S}_i \cdot \vec{S}_{i+1})^2 + \frac{1}{3} \right) \,,
\end{equation}
with spin $s=1$ at each site, and here we consider periodic boundary conditions $\vec{S}_{n+1} = \vec{S}_1$. The ground state has energy $E=0$, and $\mathbb{S}^z$-eigenvalue $M=0$; the latter implies $k=n$, 
see below \eqref{quditsup}. This state has the form \eqref{quditsup}, where the amplitudes are given (up to an overall normalization factor) by a simple matrix product state (MPS) representation
\begin{equation}
    a_{\vec{m}} = \tr \left(A^{m_n} \ldots A^{m_1} \right) \,, \qquad
    A^0 = \begin{pmatrix}
        0 & 1\\
        0 & 0
    \end{pmatrix} \,, \quad
    A^1 = \frac{1}{\sqrt{2}}\begin{pmatrix}
        -1 & 0\\
        0 & 1
    \end{pmatrix} \,, \quad
    A^2 = \begin{pmatrix}
        0 & 0\\
        -1 & 0
    \end{pmatrix} \,.
\label{AKLT-MPS}
\end{equation}
Note that all the amplitudes are real.
We have explicitly verified that the circuit \eqref{genalgorithm} with $k=n$ and $s=1$ corresponding to the amplitudes \eqref{AKLT-MPS} indeed prepares the exact AKLT ground state, for various values of $n$, see \cite{GitHubRN}.
The MPS representation of these states \eqref{AKLT-MPS} can be directly exploited to prepare these states more efficiently \cite{Smith:2022nbd, Kumaran:2024atr}.

\section{Spin-$s$ Dicke states}\label{sec:Dicke}

For given values of $(n,k,s)$, the states \eqref{quditsup} with real amplitudes 
\begin{equation}
 a_{\vec{m}} =   \sqrt{\frac{\binom{2s}{m_1} \binom{2s}{m_2} \cdots \binom{2s}{m_n}}{\binom{2sn}{k}}}
\label{Dicke-amplitudes}
\end{equation}
are so-called $su(2)$ spin-$s$ Dicke states $|D^{(s)}_{n,k}\rangle$ \cite{Nepomechie:2024fhm, Raveh:2024sku}. These states are ground states of the Hamiltonian
\begin{equation}
    H = - \vec{\mathbb{S}}^2 = - \sum_{i,j = 1}^n \vec{S}_i \cdot \vec{S}_j \,,
    \qquad \vec{\mathbb{S}} = \sum_{i=1}^n \vec{S}_i \,,
\end{equation}
with energy $E= -s n ( s n +1)$ that is independent of $k$,
and $\mathbb{S}^z$-eigenvalue $M= s n - k$. For $s=1/2$, these states reduce to ordinary qubit Dicke states \cite{Dicke:1954zz}.

We have explicitly verified that the circuit \eqref{genalgorithm} indeed prepares the exact states $|D^{(s)}_{n,k}\rangle$, for various values of $(n,k,s)$, see \cite{GitHubRN}. The high degree of symmetry of these states can be exploited to prepare these states more efficiently \cite{Nepomechie:2024fhm, Kerzner:2025uxw}.

\section{Spin-$s$ Bethe states}\label{sec:Bethe}

A remarkable feature of the spin-1/2 XXX Hamiltonian is that it is {\em quantum integrable}; its exact eigenstates can be expressed in terms of solutions of so-called Bethe equations \cite{Bethe:1931hc, Gaudin:1983, Faddeev:1996iy}. Based on \cite{Zamolodchikov:1980ku, Kulish:1981gi, Kulish:1981bi}, quantum integrable Hamiltonians for general values of spin $s$ with periodic boundary conditions
were formulated in \cite{Babujian:1983ae}. These Hamiltonians can be expressed
in terms of the polynomials $h(x,s)$ in $x$ of degree $2s$ given by 
\begin{equation}
h(x, s) = 2 \sum_{i=1}^{2s} \left(\sum_{j=1}^i \frac{1}{j}\right) \
\prod_{\substack{l=0\\ \ne i}}^{2s} \frac{x-x_l}{x_i-x_l} + \text{const} \,, 
\qquad x_l = \tfrac{1}{2}\left[l(l+1)-2s(s+1)\right] \,.
\label{hxs}
\end{equation}
Expressions for $h(x, s)$ for small values of $s$ are displayed in Table \ref{table:h}.\footnote{We adjust the additive constant in $h(x, s)$ so that there is no additive constant in the expression for the energy \eqref{energy-spin-s}, in agreement with \cite{Crampe:2011}.}

\begin{table}[ht]
\centering
\begin{tabular}{|c|c|}
\hline
$s$ & $h(x,s)$ \\   
\hline
1/2 & $ - \frac{1}{2} + 2x$  \\[0.1 in]
1 &  $\frac{1}{2}x - \frac{1}{2} x^2  $  \\[0.1 in]
3/2 & $-\frac{3}{4} -\frac{1}{8}x + \frac{1}{27}x^2 + \frac{2}{27}x^3  $\\[0.1 in]
2 & $ -\frac{1}{2} + \frac{13}{24}x + \frac{43}{432}x^2 - \frac{5}{216}x^3 - \frac{1}{144}x^4 $ \\[0.1 in]
\hline
\end{tabular}
\caption{The polynomials $h(x,s)$ \eqref{hxs} for small values of $s$.}\label{table:h}
\end{table}

\noindent
The integrable spin-$s$ Hamiltonians with periodic boundary conditions ($\vec{S}_{n+1} = \vec{S}_1$) are given by
\begin{equation}
H(s) = \sum_{i=1}^n h(x_i, s) \,, \qquad x_i = \vec{S}_i \cdot \vec{S}_{i+1} \,.
\label{Hamiltonian-spin-s}    
\end{equation}

An expression for the Bethe states (i.e., exact eigenstates of $H(s)$ in terms of Bethe roots, known as the coordinate Bethe ansatz) has been worked out for $s=1$ in \cite{Lima:1999}, and for general values of $s$ in \cite{Crampe:2011}. In our notation, the result from \cite{Crampe:2011} is given by
\begin{equation}
|\psi\rangle = \sum_{1\le x_1 \le x_2 \le \ldots \le x_k \le n} a(x_1, x_2,  \ldots, x_k)\, |x_1, x_2, \ldots, x_k \rangle\!\rangle \,,
\label{CBA-psi}
\end{equation}
where
\begin{equation}
    |x_1, x_2, \ldots, x_k \rangle\!\rangle = e^-_{x_1}\,  e^-_{x_2}\, \ldots  e^-_{x_k}\, |0\rangle^{\otimes n} \,, \qquad e^- = \sum_{j=0}^{2s-1} \sqrt{\frac{2s-j}{j+1}}\, |j+1\rangle \langle j | \,.
\label{CBA-x}
\end{equation}
Note that no more than $2s$ of the $x_i$'s can be equal, since $\left(e^-\right)^{2s+1}=0$.
Furthermore,\footnote{We correct here a typo in the expression for $A_P(\vec{\kk})$ in \cite{Crampe:2011}, namely, in the sign in front of $\frac{1}{2s}$.}
\begin{equation}
 a(x_1, x_2, \ldots, x_k) = \sum_{P \in \mathfrak{S}_k} A_P(\vec{\kk})\, e^{i \sum_{j=1}^k\kk_{Pj} x_j} \,, \quad A_P(\vec{\kk}) = \prod_{j < l}
\left(1 - \frac{1}{2s}\frac{(e^{i \kk_{Pj}}-1)(e^{i \kk_{Pl}}-1)}
{e^{i \kk_{Pj}}-e^{i \kk_{Pl}}} \right) \,,
\label{CBA-a}
\end{equation}
where $\mathfrak{S}_k$ is the group of permutations of $1, \ldots, k$. The so-called Bethe roots $\vec{\kk} = (\kk_1, \ldots, \kk_k)$ are related to $\vec{u} = (u_1, \ldots, u_k)$ by
\begin{equation}
    e^{i \kk_j} = \frac{u_j+is}{u_j-is} \,, \qquad j = 1, \ldots, k \,;
\end{equation}
and the latter are solutions of the Bethe equations
\begin{equation}
\left(\frac{u_j + i s}{u_j - i s} \right)^n = \prod_{\substack{l=1\\ \ne j}}^{k} \frac{u_j - u_l + i}{u_j - u_l - i} \,, \qquad j = 1, \ldots, k \,.
\label{BetheEqs} 
\end{equation}
Solutions of these equations for small values of $(n,k,s)$ are reported in \cite{Hao:2013rza}.
The energies are given by
\begin{equation}
    E=-\sum_{j=1}^k \frac{2s}{u_j^2 + s^2} \,,
\label{energy-spin-s}
\end{equation}
and the $\mathbb{S}^z$-eigenvalues are again given by $M= s n - k$.

Let us now recast the result \eqref{CBA-psi} into the form \eqref{quditsup}. To this end, we observe that
\begin{equation}
|x_1, x_2, \ldots, x_k \rangle\!\rangle = \prod_{j=1}^n \sqrt{\binom{2s}{\mu(j)}}\, |\mu(n) \ldots \mu(1) \rangle \,,
\end{equation}
where $\mu(j) \in \{0, 1, \ldots 2s\}$ is defined as the number of times that $j$ appears in $(x_1, x_2, \ldots, x_k)$, where $j \in \{1, 2, \ldots, n\}$. Conversely, 
\begin{equation}
|m_n \ldots m_2\, m_1 \rangle = \prod_{j=1}^n \frac{1}{\sqrt{\binom{2s}{m_j}}}\, | \underbrace{1, \ldots, 1}_{m_1}, \underbrace{2, \ldots, 2}_{m_2}, \ldots, \underbrace{n, \ldots, n}_{m_n} \rangle\!\rangle \,,
\end{equation}
where $m_1 + m_2 + \ldots + m_n = k$. 
It follows that the spin-$s$ Bethe states are given by \eqref{quditsup}, with the amplitudes given (up to an overall normalization factor) by
\begin{equation}
    a_{\vec{m}} = a(x_1, x_2, \ldots, x_k)\, \prod_{j=1}^n \sqrt{\binom{2s}{m_j}} \,,
\label{Bethe-amplitudes}
\end{equation}
where $(x_1, x_2, \ldots, x_k)=(\underbrace{1, \ldots, 1}_{m_1}, \underbrace{2, \ldots, 2}_{m_2}, \ldots, \underbrace{n, \ldots, n}_{m_n})$, and $a(x_1, x_2, \ldots, x_k)$ is given by \eqref{CBA-a}. Note that  the amplitudes depend on the Bethe roots $\vec{\kk}$ as well as on $\vec{m}$, and are generally complex-valued.

Using Bethe roots from \cite{Hao:2013rza},
we have explicitly verified that the circuit \eqref{genalgorithm} indeed prepares exact eigenstates of the Hamiltonain $H(s)$ \eqref{Hamiltonian-spin-s}, with corresponding eigenvalue \eqref{energy-spin-s}, for various values of $(n,k,s)$, see \cite{GitHubRN}.

\section{Discussion}\label{sec:Discussion}

We have formulated a deterministic algorithm for preparing a general $\mathbb{S}^z$-eigenstate of a spin-$s$ chain of length $n$ \eqref{quditsup}. These states consist of linear combinations of computational basis states $|\vec{m}\rangle$ of $n$ qudits, each with $(2s+1)$ levels, whose ditstrings $\vec{m}$ have a fixed digit sum. Exploiting a Gray code for bounded integer compositions \cite{Walsh:2000}, whose consecutive ditstrings obey the Gray property \eqref{Gray}, the quantum state is prepared by applying corresponding Gray gates \eqref{genalgorithm}. For $s=1/2$, this algorithm reduces to the one in \cite{Farias:2024ejh}. 

We have shown how to use this algorithm to prepare spin-$s$ Bethe states, which heretofore had been accomplished only for $s=1/2$. These Bethe states can potentially be used as trial states in a VQE setup \cite{Tilly:2021jem}, treating the Bethe roots $\vec{\kk}$ as variational parameters, in order to estimate the Bethe roots, as has been done for $s=1/2$ \cite{Nepomechie:2020, Raveh:2024cfh}. It should be possible to generalize the coordinate Bethe ansatz \cite{Crampe:2011} to the integrable anisotropic (XXZ) case with periodic boundary conditions, as well as to the case of integrable $U(1)$-invariant open boundary conditions, to which our algorithm could also be applied.
It should also be possible to use this algorithm to prepare exact eigenstates of yet other $U(1)$-invariant integrable models, such as the $A_2^{(2)}$ spin chain \cite{Izergin:1980pe, Lima:1999}. We emphasize, however, that 
this way of preparing Bethe states has a high circuit complexity, both quantum (circuit size and depth) and classical (the computation of the amplitudes entails summation over all permutations \eqref{CBA-a}). It would be very interesting if the integrable structure of Bethe states could be exploited to prepare them more efficiently, as is possible for the AKLT ground state and Dicke states due to their special features. We have restricted here to eigenstates of models with a single $U(1)$ symmetry; it may also be interesting to generalize this work to models with more than one $U(1)$ symmetry.

\section*{Acknowledgements}
We thank Nicolas Cramp\'e for helpful correspondence.
RN is supported in part by the National Science Foundation under grant PHY 2310594, and by a Cooper fellowship.  

\appendix

\section{Walsh's Gray code for bounded compositions}\label{sec:Walsh}

To make this paper self-contained, we review here the non-recursive $\mathcal{O}(n)$ Gray code for bounded compositions that is given in Section 2 of \cite{Walsh:2000}.\footnote{A more sophisticated
``loop-free'' ($\mathcal{O}(1)$) implementation is given in Section 3 of \cite{Walsh:2000}.}
Following \cite{Walsh:2000}, here we denote the ditstrings by $\vec{g} = (g_1, \ldots, g_n)$; and we look for a list of bounded compositions of $k$ 
\begin{equation}
g_1 + \ldots + g_n = k\,, \qquad 0 \le g_i \le 2s  
\quad \text{for all}\quad i \,,
\label{constraints}
\end{equation}
with the Gray property.
A ``part'' $g_i$ of $\vec{g}$ has the ``prefix'' $(g_1, \ldots, g_{i-1})$ and the ``suffix'' $(g_{i+1}, \ldots, g_n)$. For each part $g_i$, we define the ``suffix sum''  $S_i$ and the ``prefix capacity'' $M_i$ by
\begin{equation}
    S_i = \sum_{l=i+1}^n g_l \,, \qquad M_i = 2 s (i-1) \,,
\end{equation}
respectively. The combined constraints \eqref{constraints} imply that each part  $g_i$ must in fact satisfy
\begin{equation}
L_i \le g_i \le U_i \,, \quad L_i =\max(0, k -S_i -M_i)\,, 
\quad U_i = \min( 2s, k -S_i) \,.
\label{LU}
\end{equation}
If $S_i$ is {\em even}, we refer to $L_i$ and $U_i$ as the {\bf first} and {\bf last} values of $g_i$, respectively.

\noindent
If $S_i$ is {\em odd}, we refer to $U_i$ and $L_i$ as the {\bf first} and {\bf last} values of $g_i$, respectively.

The initial composition is the lexicographically largest one.

To find the successor of the ``current'' composition $(g_1, \ldots, g_n)$:
\begin{enumerate}
\item  Find the first pivot ($i$): scan from $i=2$ to $n$ to find the the smallest index $i$ such that $g_i$ is {\em not} at its  {\bf last} value relative to its $S_i$.
If no such $i$ exists, the current composition is the final one in the list.
\item Change 
$g_i \mapsto g_i + (-1)^{S_i} $; that is, raise (lower) $g_i$ by 1 if $S_i$ is even (odd), respectively.

\item  Find the second pivot ($j$): scan the prefix of $g_i$, from $j=i-1$ down to 1, to find the largest $j<i$ such that $g_j$ is {\em not} at its {\em new} {\bf first} value relative to its updated $S_j$.
\item Change  $g_j$  to its new {\bf first} value, which either lowers or raises it by 1.
\end{enumerate}

An example for the case $(n,k,s)=(3,3,1)$ is displayed in Table \ref{table:WalshGrayExample}. The initial ($l=0$) composition is the lexicographically largest one, while the final  ($l=6$) composition is the lexicographically
smallest one. See also \cite{GitHubRN}.

\begin{table}[ht]
\centering
\begin{tabular}{|c|c|}
\hline
$l$ & $(g_1, g_2, g_3)$  \\   
\hline
0 & (\uuline{2},\uline{1}, 0)  \\
1 & (\uuline{1}, 2, \uline{0})  \\
2 & (\uuline{0}, \uline{2}, 1) \\
3 & (\uuline{1}, \uline{1}, 1) \\
4 & (\uuline{2}, 0, \uline{1}) \\
5 & (\uuline{1}, \uline{0}, 2) \\
6 & (0, 1, 2) \\
\hline
\end{tabular}
\caption{Walsh's Gray code for $(n,k,s) = (3, 3, 1)$. The parts $g_i$ at the first and second pivots are designated with a single and double underline, respectively.}
\label{table:WalshGrayExample}
\end{table}


\begin{thebibliography}{10}

\bibitem{Bartschi2019}
A.~B\"artschi and S.~Eidenbenz, ``{Deterministic preparation of Dicke states},'' {\em Lecture Notes in Computer Science} (2019) 126--139, \href{http://arxiv.org/abs/1904.07358}{{\ttfamily arXiv:1904.07358 [quant-ph]}}.

\bibitem{Wang:2021}
Y.~Wang and B.~M. Terhal, ``{Preparing Dicke states in a spin ensemble using phase estimation},'' {\em Phys. Rev. A} {\bfseries 104} no.~3, (2021) , \href{http://arxiv.org/abs/2104.14310}{{\ttfamily arXiv:2104.14310 [quant-ph]}}.

\bibitem{Buhrman:2023rft}
H.~Buhrman, M.~Folkertsma, B.~Loff, and N.~M.~P. Neumann, ``{State preparation by shallow circuits using feed forward},'' \href{http://dx.doi.org/10.22331/q-2024-12-09-1552}{{\em Quantum} {\bfseries 8} (2024) 1552}, \href{http://arxiv.org/abs/2307.14840}{{\ttfamily arXiv:2307.14840 [quant-ph]}}.

\bibitem{Bond:2023yry}
L.~J. Bond, M.~J. Davis, J.~Min{\'a}{\v{r}}, R.~Gerritsma, G.~K. Brennen, and A.~Safavi-Naini, ``{Global variational quantum circuits for arbitrary symmetric state preparation},'' \href{http://dx.doi.org/10.1103/PhysRevResearch.7.L022072}{{\em Phys. Rev. Res.} {\bfseries 7} no.~2, (2025) L022072}, \href{http://arxiv.org/abs/2312.05060}{{\ttfamily arXiv:2312.05060 [quant-ph]}}.

\bibitem{Piroli:2024ckr}
L.~Piroli, G.~Styliaris, and J.~I. Cirac, ``{Approximating Many-Body Quantum States with Quantum Circuits and Measurements},'' \href{http://dx.doi.org/10.1103/PhysRevLett.133.230401}{{\em Phys. Rev. Lett.} {\bfseries 133} no.~23, (2024) 230401}, \href{http://arxiv.org/abs/2403.07604}{{\ttfamily arXiv:2403.07604 [quant-ph]}}.

\bibitem{Yu:2024szp}
J.~Yu, S.~R. Muleady, Y.-X. Wang, N.~Schine, A.~V. Gorshkov, and A.~M. Childs, ``{Efficient Preparation of Dicke States},'' \href{http://dx.doi.org/10.1103/9gjk-rgql}{{\em Phys. Rev. Lett.} {\bfseries 136} no.~3, (2026) 030601}, \href{http://arxiv.org/abs/2411.03428}{{\ttfamily arXiv:2411.03428 [quant-ph]}}.

\bibitem{Liu:2024taj}
Z.~Liu, A.~M. Childs, and D.~Gottesman, ``{Low-depth quantum symmetrization},'' \href{http://arxiv.org/abs/2411.04019}{{\ttfamily arXiv:2411.04019 [quant-ph]}}.

\bibitem{Vasconcelos:2026sqs}
F.~Vasconcelos and M.~R. Joshi, ``{Constant-Depth Unitary Preparation of Dicke States},'' \href{http://arxiv.org/abs/2601.10693}{{\ttfamily arXiv:2601.10693 [quant-ph]}}.

\bibitem{VanDyke:2021kvq}
J.~S. Van~Dyke, G.~S. Barron, N.~J. Mayhall, E.~Barnes, and S.~E. Economou, ``{Preparing Bethe Ansatz Eigenstates on a Quantum Computer},'' {\em PRX Quantum} {\bfseries 2} (2021) 040329, \href{http://arxiv.org/abs/2103.13388}{{\ttfamily arXiv:2103.13388 [quant-ph]}}.

\bibitem{VanDyke:2021nuz}
J.~S. Van~Dyke, E.~Barnes, S.~E. Economou, and R.~I. Nepomechie, ``{Preparing exact eigenstates of the open XXZ chain on a quantum computer},'' {\em J. Phys. A} {\bfseries 55} no.~5, (2022) 055301, \href{http://arxiv.org/abs/2109.05607}{{\ttfamily arXiv:2109.05607 [quant-ph]}}.

\bibitem{Li:2022czv}
W.~Li, M.~Okyay, and R.~I. Nepomechie, ``{Bethe states on a quantum computer: success probability and correlation functions},'' {\em J. Phys. A} {\bfseries 55} no.~35, (2022) 355305, \href{http://arxiv.org/abs/2201.03021}{{\ttfamily arXiv:2201.03021 [quant-ph]}}.

\bibitem{Sopena:2022ntq}
A.~Sopena, M.~H. Gordon, D.~Garc\'\i{}a-Mart\'\i{}n, G.~Sierra, and E.~L\'opez, ``{Algebraic Bethe Circuits},'' {\em Quantum} {\bfseries 6} (2022) 796, \href{http://arxiv.org/abs/2202.04673}{{\ttfamily arXiv:2202.04673 [quant-ph]}}.

\bibitem{Ruiz:2023rew}
R.~Ruiz, A.~Sopena, M.~H. Gordon, G.~Sierra, and E.~L\'opez, ``{The Bethe Ansatz as a Quantum Circuit},'' \href{http://dx.doi.org/10.22331/q-2024-05-23-1356}{{\em Quantum} {\bfseries 8} (2024) 1356}, \href{http://arxiv.org/abs/2309.14430}{{\ttfamily arXiv:2309.14430 [quant-ph]}}.

\bibitem{Ruiz:2025qmt}
R.~Ruiz, A.~Sopena, E.~L{\'o}pez, G.~Sierra, and B.~Pozsgay, ``{Bethe Ansatz, quantum circuits, and the F-basis},'' \href{http://dx.doi.org/10.21468/SciPostPhys.18.6.187}{{\em SciPost Phys.} {\bfseries 18} no.~6, (2025) 187}, \href{http://arxiv.org/abs/2411.02519}{{\ttfamily arXiv:2411.02519 [quant-ph]}}.

\bibitem{Sahu:2024toz}
S.~Sahu and G.~Vidal, ``{Fractal decompositions and tensor network representations of Bethe wavefunctions},'' \href{http://dx.doi.org/10.21468/SciPostPhysCore.8.4.067}{{\em SciPost Phys. Core} {\bfseries 8} (2025) 067}, \href{http://arxiv.org/abs/2412.00923}{{\ttfamily arXiv:2412.00923 [quant-ph]}}.

\bibitem{Yeo:2025tph}
H.~Yeo, H.~E. Kim, I.~Sohn, and K.~Jeong, ``{Reducing circuit depth in quantum state preparation for quantum simulation using measurements and feedforward},'' \href{http://dx.doi.org/10.1103/PhysRevApplied.23.054066}{{\em Phys. Rev. Applied} {\bfseries 23} no.~5, (2025) 054066}, \href{http://arxiv.org/abs/2501.02929}{{\ttfamily arXiv:2501.02929 [quant-ph]}}.

\bibitem{Raveh:2024llj}
D.~Raveh and R.~I. Nepomechie, ``{Deterministic Bethe state preparation},'' \href{http://dx.doi.org/10.22331/q-2024-10-24-1510}{{\em Quantum} {\bfseries 8} (2024) 1510}, \href{http://arxiv.org/abs/2403.03283}{{\ttfamily arXiv:2403.03283 [quant-ph]}}.

\bibitem{Farias:2024ejh}
R.~M.~S. Farias, T.~O. Maciel, G.~Camilo, R.~Lin, S.~Ramos-Calderer, and L.~Aolita, ``{Quantum encoder for fixed-Hamming-weight subspaces},'' \href{http://dx.doi.org/10.1103/PhysRevApplied.23.044014}{{\em Phys. Rev. Applied} {\bfseries 23} no.~4, (2025) 044014}, \href{http://arxiv.org/abs/2405.20408}{{\ttfamily arXiv:2405.20408 [quant-ph]}}.

\bibitem{Mao:2024hfg}
R.~Mao, G.~Tian, and X.~Sun, ``{Toward optimal circuit size for sparse quantum state preparation},'' \href{http://dx.doi.org/10.1103/PhysRevA.110.032439}{{\em Phys. Rev. A} {\bfseries 110} no.~3, (2024) 032439}, \href{http://arxiv.org/abs/2404.05147}{{\ttfamily arXiv:2404.05147 [quant-ph]}}.

\bibitem{Li:2025zzw}
Y.~Li, G.~Tian, X.~He, and X.~Sun, ``{Preparation of Hamming-Weight-Preserving Quantum States with Log-Depth Quantum Circuits},'' \href{http://arxiv.org/abs/2508.14470}{{\ttfamily arXiv:2508.14470 [quant-ph]}}.

\bibitem{Luo:2025zjt}
J.~Luo and L.~Li, ``{Optimal Circuit Size for Fixed-Hamming-Weight Quantum States Preparation},'' \href{http://arxiv.org/abs/2508.17197}{{\ttfamily arXiv:2508.17197 [quant-ph]}}.

\bibitem{Smith:2022nbd}
K.~C. Smith, E.~Crane, N.~Wiebe, and S.~M. Girvin, ``{Deterministic Constant-Depth Preparation of the AKLT State on a Quantum Processor Using Fusion Measurements},'' \href{http://dx.doi.org/10.1103/PRXQuantum.4.020315}{{\em PRX Quantum} {\bfseries 4} no.~2, (2023) 020315}, \href{http://arxiv.org/abs/2210.17548}{{\ttfamily arXiv:2210.17548 [quant-ph]}}.

\bibitem{Kumaran:2024atr}
K.~Kumaran, F.~Alam, N.~Eassa, K.~Ferris, X.~Xiao, L.~Cincio, N.~Bronn, and A.~Banerjee, ``{Transmon qutrit-based simulation of spin-1 AKLT systems},'' \href{http://arxiv.org/abs/2412.19786}{{\ttfamily arXiv:2412.19786 [quant-ph]}}.

\bibitem{Nepomechie:2024fhm}
R.~I. Nepomechie, F.~Ravanini, and D.~Raveh, ``{Spin-s Dicke states and their preparation},'' \href{http://dx.doi.org/10.1002/qute.202400057}{{\em Adv. Quantum Technol.} {\bfseries 7} no.~12, (2024) 2400057}, \href{http://arxiv.org/abs/2402.03233}{{\ttfamily arXiv:2402.03233 [quant-ph]}}.

\bibitem{Raveh:2024sku}
D.~Raveh and R.~I. Nepomechie, ``{Dicke states as matrix product states},'' \href{http://dx.doi.org/10.1103/PhysRevA.110.052438}{{\em Phys. Rev. A} {\bfseries 110} no.~5, (2024) 052438}, \href{http://arxiv.org/abs/2408.04729}{{\ttfamily arXiv:2408.04729 [quant-ph]}}.

\bibitem{Kerzner:2025uxw}
N.~B. Kerzner, F.~Galeazzi, and R.~I. Nepomechie, ``{Simple ways of preparing qudit Dicke states},'' \href{http://dx.doi.org/10.2478/qic-2025-0036}{{\em Quant. Inf. Comp.} {\bfseries 25} (2025) 668–686}, \href{http://arxiv.org/abs/2507.13308}{{\ttfamily arXiv:2507.13308 [quant-ph]}}.

\bibitem{Bethe:1931hc}
H.~Bethe, ``{On the theory of metals. 1. Eigenvalues and eigenfunctions for the linear atomic chain},'' {\em Z. Phys.} {\bfseries 71} (1931) 205--226.

\bibitem{Gaudin:1983}
M.~Gaudin, {\em {La fonction d'onde de Bethe}}.
\newblock Masson, 1983.
\newblock English translation by J.-S. Caux, {\em The Bethe wavefunction}, CUP, 2014.

\bibitem{Faddeev:1996iy}
L.~D. Faddeev, ``{How algebraic Bethe ansatz works for integrable models},'' in {\em Sym\'etries Quantiques (Les Houches Summer School Proceedings vol 64)}, A.~Connes, K.~Gawedzki, and J.~Zinn-Justin, eds., pp.~149--219.
\newblock North Holland, 1998.
\newblock \href{http://arxiv.org/abs/hep-th/9605187}{{\ttfamily arXiv:hep-th/9605187 [hep-th]}}.

\bibitem{Wilf:1989}
H.~S. Wilf, {\em {Combinatorial algorithms:an update}}.
\newblock SIAM, 1989.

\bibitem{savage:1997}
C.~Savage, ``{A survey of combinatorial Gray codes},'' {\em SIAM Rev.} {\bfseries 39} (1997) 605–629.

\bibitem{mutze:2012}
T.~M{\"u}tze, ``{Combinatorial Gray codes—an updated survey},'' {\em Elect. J. Combinatorics} (2012) DS26--Sep, \href{http://arxiv.org/abs/2202.01280}{{\ttfamily arXiv:2202.01280 [math.CO]}}.

\bibitem{Dicke:1954zz}
R.~H. Dicke, ``{Coherence in Spontaneous Radiation Processes},'' {\em Phys. Rev.} {\bfseries 93} (1954) 99--110.

\bibitem{Affleck:1987vf}
I.~Affleck, T.~Kennedy, E.~H. Lieb, and H.~Tasaki, ``{Rigorous Results on Valence Bond Ground States in Antiferromagnets},'' \href{http://dx.doi.org/10.1103/PhysRevLett.59.799}{{\em Phys. Rev. Lett.} {\bfseries 59} (1987) 799}.

\bibitem{Affleck:1987cy}
I.~Affleck, T.~Kennedy, E.~H. Lieb, and H.~Tasaki, ``{Valence Bond Ground States in Isotropic Quantum Antiferromagnets},'' \href{http://dx.doi.org/10.1007/BF01218021}{{\em Commun. Math. Phys.} {\bfseries 115} (1988) 477}.

\bibitem{Zamolodchikov:1980ku}
A.~B. Zamolodchikov and V.~A. Fateev, ``{Model factorized S matrix and an integrable Heisenberg chain with spin 1},'' {\em Sov. J. Nucl. Phys.} {\bfseries 32} (1980) 298--303.

\bibitem{Kulish:1981gi}
P.~P. Kulish, N.~{\relax Yu}. Reshetikhin, and E.~K. Sklyanin, ``{Yang-Baxter Equation and Representation Theory. 1.},''
\href{http://dx.doi.org/10.1007/BF02285311}{{\em Lett. Math. Phys.} {\bfseries 5} (1981) 393--403}.

\bibitem{Kulish:1981bi}
P.~P. Kulish and E.~K. Sklyanin, ``{Quantum spectral transform method. Recent developments},''
{\em Lect. Notes Phys.} {\bfseries 151} (1982) 61--119.

\bibitem{Babujian:1983ae}
H.~M. Babujian, ``{Exact solution of the isotropic Heisenberg chain with arbitary spins: thermodynamics of the model},'' \href{http://dx.doi.org/10.1016/0550-3213(83)90668-5}{{\em Nucl. Phys. B} {\bfseries 215} (1983) 317--336}.

\bibitem{Lima:1999}
A.~{Lima-Santos}, ``{Bethe ans{\"a}tze for 19-vertex models},'' {\em J. Phy. A} {\bfseries 32} no.~10, (1999) 1819--1839, \href{http://arxiv.org/abs/hep-th/9807219}{{\ttfamily arXiv:hep-th/9807219 [hep-th]}}.

\bibitem{Crampe:2011}
N.~{Cramp{\'e}}, E.~{Ragoucy}, and L.~{Alonzi}, ``{Coordinate Bethe Ansatz for Spin s XXX Model},'' {\em SIGMA} {\bfseries 7} (2011) 006, \href{http://arxiv.org/abs/1009.0408}{{\ttfamily arXiv:1009.0408 [math-ph]}}.

\bibitem{cirq}
{Cirq Developers}, ``Cirq,'' 2025.
\newblock \url{https://doi.org/10.5281/zenodo.4062499}.

\bibitem{GitHubRN}
\url{https://github.com/nepomechie/spin-s-U1-eigenstate-preparation}.

\bibitem{Walsh:2000}
T.~R. Walsh, ``{Loop-free sequencing of bounded integer compositions},'' {\em J. Comb. Math. Comb. Comp.} {\bfseries 33} (2000) 323--345.

\bibitem{Klingsberg:1982}
P.~Klingsberg, ``{A Gray code for compositions},'' {\em J. Algorithms} {\bfseries 3} (1982) 41--44.

\bibitem{Even:1973}
S.~Even, {\em {Algorithmic Combinatorics}}.
\newblock Macmillan, 1973.

\bibitem{horan:2014}
V.~Horan and G.~Hurlbert, ``{Gray codes and overlap cycles for restricted weight words},'' {\em Discrete Math. Alg. Appl.} {\bfseries 06} (2014) 1450062, \href{http://arxiv.org/abs/1403.1818}{{\ttfamily arXiv:1403.1818 [math.CO]}}.

\bibitem{Wang:2020}
Y.~{Wang}, Z.~{Hu}, B.~C. {Sanders}, and S.~{Kais}, ``{Qudits and high-dimensional quantum computing},'' {\em Front. Phys.} {\bfseries 8} (2020) 479, \href{http://arxiv.org/abs/2008.00959}{{\ttfamily arXiv:2008.00959 [quant-ph]}}.

\bibitem{Zi:2023drn}
W.~Zi, Q.~Li, and X.~Sun, \href{http://dx.doi.org/10.1109/DAC56929.2023.10247925}{``{Optimal Synthesis of Multi-Controlled Qudit Gates},''} in {\em 60th ACM/IEEE Design Automation Conference (DAC), San Francisco, CA, USA}, pp.~1--6.
\newblock 2023.
\newblock \href{http://arxiv.org/abs/2303.12979}{{\ttfamily arXiv:2303.12979 [quant-ph]}}.

\bibitem{Hao:2013rza}
W.~Hao, R.~I. Nepomechie, and A.~J. Sommese, ``{Singular solutions, repeated roots and completeness for higher-spin chains},'' \href{http://dx.doi.org/10.1088/1742-5468/2014/03/P03024}{{\em J. Stat. Mech.} {\bfseries 1403} (2014) P03024}, \href{http://arxiv.org/abs/1312.2982}{{\ttfamily arXiv:1312.2982 [math-ph]}}.

\bibitem{Tilly:2021jem}
J.~Tilly {\em et~al.}, ``{The Variational Quantum Eigensolver: A review of methods and best practices},'' \href{http://dx.doi.org/10.1016/j.physrep.2022.08.003}{{\em Phys. Rept.} {\bfseries 986} (2022) 1--128}, \href{http://arxiv.org/abs/2111.05176}{{\ttfamily arXiv:2111.05176 [quant-ph]}}.

\bibitem{Nepomechie:2020}
R.~I. Nepomechie, ``{Bethe ansatz on a quantum computer?},'' {\em Quantum Inf. Comp.} {\bfseries 21} (2021) 255--265, \href{http://arxiv.org/abs/2010.01609}{{\ttfamily arXiv:2010.01609 [quant-ph]}}.

\bibitem{Raveh:2024cfh}
D.~Raveh and R.~I. Nepomechie, ``{Estimating Bethe roots with VQE},'' \href{http://dx.doi.org/10.1088/1751-8121/ad6db2}{{\em J. Phys. A} {\bfseries 57} no.~35, (2024) 355303}, \href{http://arxiv.org/abs/2404.18244}{{\ttfamily arXiv:2404.18244 [quant-ph]}}.

\bibitem{Izergin:1980pe}
A.~G. Izergin and V.~E. Korepin, ``{The inverse scattering method approach to the quantum Shabat-Mikhailov model},''
\href{http://dx.doi.org/10.1007/BF01208496}{{\em Commun. Math. Phys.} {\bfseries 79} (1981) 303}.

\end{thebibliography}

\providecommand{\href}[2]{#2}\begingroup\raggedright\endgroup

\end{document}